\documentclass[sigconf, nonacm]{acmart}

\usepackage{xspace}
\usepackage{listings}

\colorlet{blue}{black}
\colorlet{orange}{black}
\colorlet{red}{black}

\usepackage{balance}
\usepackage[labelformat=simple]{subcaption}

\usepackage[nolist,printonlyused]{acronym}
\usepackage{tcolorbox}

\newcommand{\cnumber}[1]{\tikz[baseline=-0.6ex]{\node[circle, draw=black, inner sep=0pt, minimum size=0.9em, fill=white, text=black] {\textbf{{\footnotesize {#1}}}};}}

\newcommand{\sysname}{\textit{Poodle}\xspace}

\newenvironment{takeaway}
  {\par\noindent\begingroup\bfseries Takeaway: \itshape\ignorespaces}
  {\endgroup\par}

\usepackage{todonotes}

\newcommand\vldbyear{2026}
\newcommand\vldbworkshop{Novel Optimizations for Visionary AI Systems (NOVAS)}
\newcommand\vldbauthors{\authors}
\newcommand\vldbtitle{\shorttitle}
\newcommand\vldbavailabilityurl{https://github.com/hpides/poodle}
\newcommand\vldbpagestyle{plain}

\begin{acronym}
    \acro{llm}[LLM]{large language model}
    \acro{ml}[ML]{machine learning}
    \acro{dl}[DL]{deep learning}
    \acro{ai}[AI]{artificial intelligence}
    \acro{nn}[NN]{neural network}
    \acro{hf}[HF]{Hugging Face}
    \acro{jitr}[JITR]{just-in-time model replacement}
    \acro{surrogatemodel}[sur\-ro\-gate mo\-del]{sur\-ro\-gate mo\-del}
    \acro{rectask}[recurring task]{recurring task}
    \acro{oom}[OOM]{order of magnitude}
    \acrodefplural{oom}[OOMs]{orders of magnitude}
\end{acronym}

\acused{surrogatemodel}
\acused{rectask}

\begin{document}
\title{Poodle: Seamlessly Scaling Down Large Language Models with Just-in-Time Model Replacement}

\author{Nils Strassenburg}
\email{nils.strassenburg@hpi.de}
\affiliation{%
  \institution{Hasso Plattner Institute, Uni Potsdam}
}

\author{Boris Glavic}
\email{bglavic@uic.edu}
\affiliation{%
  \institution{University of Illinois Chicago}
}

\author{Tilmann Rabl}
\email{tilmann.rabl@hpi.de}
\affiliation{%
  \institution{Hasso Plattner Institute, Uni Potsdam}
}

\begin{abstract}
Businesses increasingly rely on \acp{llm} to automate simple repetitive tasks instead of developing custom machine learning models. \acp{llm} require few, if any, training examples and can be utilized by users without expertise in model development.
However, this comes at the cost of substantially higher resource and energy consumption compared to smaller models, which often achieve similar predictive performance for simple tasks.
In this paper, we present our vision for \emph{\ac{jitr}}, where, upon identifying a \acs{rectask} in calls to an \ac{llm}, the model is replaced transparently with a cheaper alternative that performs well for this specific task.
\Ac{jitr} retains the ease of use and low development effort of \acp{llm}, while saving significant cost and energy.
We discuss the main challenges in realizing our vision regarding the identification of \acsp{rectask} and the creation of a custom model.
Specifically, we argue that \emph{model search} and transfer learning will play a crucial role in \ac{jitr} to efficiently identify and fine-tune models for a \acs{rectask}.
\color{blue}
Using our \ac{jitr} prototype \sysname, we reduce inference time by up to 7.5$\times$ compared to a self-hosted LLM and save more than \$2,200 per 1M requests compared to a flagship hosted LLM, while achieving accuracy competitive with the LLM baseline.
\color{black}


\end{abstract}

\maketitle

\pagestyle{\vldbpagestyle}
\begingroup\small\noindent\raggedright\textbf{VLDB Workshop Reference Format:}\\
\vldbauthors. \vldbtitle. VLDB \vldbyear\ Workshop: \vldbworkshop.\\ 
\endgroup
\begingroup
\renewcommand\thefootnote{}\footnote{\noindent
This work is licensed under the Creative Commons BY-NC-ND 4.0 International License. Visit \url{https://creativecommons.org/licenses/by-nc-nd/4.0/} to view a copy of this license. For any use beyond those covered by this license, obtain permission by emailing \href{mailto:info@vldb.org}{info@vldb.org}. Copyright is held by the owner/author(s). Publication rights licensed to the VLDB Endowment. \\
\raggedright Proceedings of the VLDB Endowment. 
ISSN 2150-8097. \\
}\addtocounter{footnote}{-1}\endgroup

\ifdefempty{\vldbavailabilityurl}{}{
\vspace{.3cm}
\begingroup\small\noindent\raggedright\textbf{VLDB Workshop Artifact Availability:}\\
The source code, data, and/or other artifacts have been made available at \url{\vldbavailabilityurl}.
\endgroup
}

\section{Introduction}
\label{sec:intro}

\acused{surrogatemodel}
\acused{rectask}

With the advent of \acfp{llm} and their availability through cloud-based services, using \ac{ml} has never been so easy.
\ac{llm} providers offer simple API integration, zero model development cost for customers, no need for extensive data collection or labeling, and instant access to state-of-the-art capabilities without requiring  AI expertise.
Organizations that previously developed custom models now use \acp{llm} to reduce development cost and time, while organizations without prior \ac{ml} experience adopt \acp{llm} for \ac{ml} automation.
As a result, companies such as Snowflake, Google, or McKinsey report that many organizations
offload simple, \aclp{rectask} such as user sentiment classification, satisfaction classification, or churn risk identification to complex \acp{llm}~\cite{mckinsey2024customercare, barkley2024forrester, chui2023ceo, Liskowski2026Cortex, Chung2026100x} even though they could be handled by much smaller, specialized models at lower cost. Furthermore, the increasing use of semantic predicates~\cite{lao2025sembench} in analytical workloads results in more repetitive and predictable \ac{llm} calls.

\begin{figure}[t]
    \centering
    \includegraphics[width=\columnwidth]{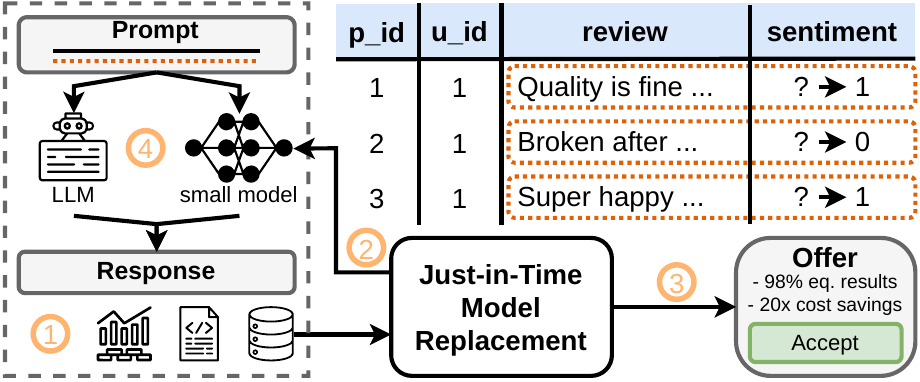}
    \caption{\textcolor{blue}{\acs{jitr} applied to sentiment classification. Reviews are sent to an \acs{llm} for sentiment classification. \acs{jitr} (1) detects the recurring task, (2) develops and monitors a surrogate model, (3) offers the user to switch to the surrogate, and (4) replaces the \acs{llm} with \acl{surrogatemodel} upon acceptance.}}
    \label{fig:running-use-case}
\end{figure}

\newcommand{\startupname}{S\xspace}

As a running example, consider the following scenario shown in \autoref{fig:running-use-case}:
Startup \startupname continuously collects product reviews in a database table.
To monitor customer satisfaction, \startupname uses a flagship \ac{llm} to predict sentiment from each review's text.
If the company has access to a database with semantic operators, this can be expressed in SQL. Otherwise, a developer is tasked with prompt engineering and writing the glue code to process each review by
calling the \ac{llm} provider's API, parsing the response, and inserting it into the sentiment column.
For simple tasks like this sentiment classification task, a smaller model would  surely suffice and reduce operational cost significantly compared to inference with an \ac{llm}.
However, developing such a model results in a longer development cycle and requires significantly more expertise and computational resources to match the \ac{llm}'s accuracy.
The developer would have to (i) generate a training dataset, (ii) find an appropriate model (e.g., on \emph{HuggingFace}~\cite{hugging_face_hugging_2025}), and fine-tune it, (iii) test and deploy it, and (iv) monitor its performance.
Thus, even though an LLM results in higher operational cost, the company still opts for this option due to significantly lower development cost.

In this paper, we present \emph{\acf{jitr}}, our vision for achieving the best of both worlds: combining the low development cost of using \acp{llm} with the low inference cost of specialized models.
Similar to \emph{just-in-time compilation}, where interpreted code is replaced with compiled code, a \ac{jitr} system replaces \acp{llm} with surrogate models in four steps as shown in \autoref{fig:running-use-case}.
\cnumber{1}~The \ac{jitr} system monitors \ac{llm} requests, identifies \emph{\aclp{rectask}}, and records request-response pairs.
\cnumber{2}~Once sufficient evidence has been observed for a \ac{rectask}, a \emph{\ac{surrogatemodel}} is generated using the recorded responses as training data.
\cnumber{3}~If the surrogate model performs well and the user approves, \cnumber{4}~future requests are handled using the \ac{surrogatemodel}, whose performance is monitored to intervene if necessary.

We envision \ac{jitr} being applied in many different scenarios.
First, \ac{llm} providers can leverage \ac{jitr} to lower serving costs and offer new pricing models. For example, upon detecting repeated instances of \startupname's sentiment analysis task, a provider could offer \startupname a price reduction for switching to a fine-tuned \emph{BERT} model.
Second, organizations self-hosting \acp{llm} may use \ac{jitr} to reduce cost and energy consumption for \acp{rectask}.
Third, LLM-augmented database engines~\cite{Chung2026100x, Liskowski2026Cortex,patel2024lotus, liu2025palimpzest, jo2024thalamusdb} that answer queries using semantic operators over structured and unstructured data can benefit from \ac{jitr}.
\ac{jitr} extends the pool of models these systems can choose from with significantly cheaper expert models.

\ac{jitr} is not the first proposal to reduce inference cost by replacing \acp{llm} with cheaper models.
\color{red}
Routing approaches ~\cite{ong2025routellmlearningroutellms} route requests among a fixed set of models, trying to select the cheapest model that still meets a target quality level. Cascading-based approaches like LOTUS~\cite{patel2024lotus} or Snowflake Cortex~\cite{Liskowski2026Cortex} route each query through a sequence of increasingly expensive models until a model gives a sufficient answer. 
Palimpzest~\cite{liu2025palimpzest} explores the Pareto front between model cost and quality to select a model, and researchers at Google propose pairing an embedding model with a cheap proxy~\cite{Chung2026100x}.
\ac{jitr}, however, is unique in being specifically designed to automatically detect and optimize \acsp{rectask}. Its task-specificity enables more specialized models to be created and used, which typically reduces inference cost significantly and often yields better accuracy than using a smaller but still generic \ac{llm}.
\color{black}

Qin et al.~\cite{qian2023creator} and Cai et al.~\cite{cai2024large} propose to automatically generate Python code for a task using an \ac{llm}.
While this works for non-semantic operations, such as extracting text from REST API responses, it fails for tasks like sentiment classification.
Given a task, Shen et al.~\cite{shen2023hugginggpt} use an \ac{llm} to coordinate task solving by finding and combining different models on HuggingFace. Models are selected solely based on their description. However, even high-quality documentation is often not sufficient for predicting the performance of a model on a new task~\cite{renggli_which_2022, li_guided_2023, achille_task2vec_2019, alsatian}.

For \ac{jitr} to be effective, several conditions have to be met:
(i) \acp{rectask} can be identified successfully from requests to an \ac{llm};
(ii) the logged requests and \ac{llm} responses provide sufficient training data for developing a \ac{surrogatemodel};
(iii) the generated \ac{surrogatemodel} has satisfactory performance;
(iv) monitoring can detect performance degradation of deployed \acp{surrogatemodel};
and (v) the total cost of task identification, model development, and monitoring can be amortized due to the reduced inference cost of the \ac{surrogatemodel}.

We identify condition (v) as the most challenging due to the high cost of developing a \ac{surrogatemodel}.
We propose to generate a \ac{surrogatemodel} by using the training data generated during task identification to first
automatically select a pre-trained base model, and afterward perform knowledge distillation~\cite{hinton2015distillingknowledgeneuralnetwork}.
The rationale for extending model development beyond traditional knowledge distillation is that, when the right base model is chosen, fine-tuning requires significantly less training data, converges faster, and achieves higher accuracy than training a new model from scratch~\cite{renggli_which_2022, renggli_shift_2022}.
This increases the likelihood that the resources spent on identifying the task and creating the \ac{surrogatemodel} will be amortized by repeated execution of the task with the \ac{surrogatemodel}.
Thus, the key to minimizing \ac{surrogatemodel} development overheads and achieving our vision of \ac{jitr} is to develop a performant model store offering effective techniques for searching base models for fine-tuning.
While existing model search approaches have been shown to be effective~\cite{renggli_which_2022, renggli_shift_2022, li_guided_2023},
further research is needed to speed up model search from a systems perspective and to scale model search to a rapidly increasing number of available base models~\cite{hugging_face_hugging_2025}.
The main contributions of this work are:
\begin{itemize}
\item We present our vision for \emph{\ac{jitr}} to reduce inference cost by transparently replacing \acp{llm} with \acp{surrogatemodel}, discuss challenges, and identify research questions (\autoref{sec:vision}).
\item We introduce \sysname, a prototype \ac{jitr} system built on top of an efficient model search system (\autoref{sec:poodle}).
\item We conduct a preliminary evaluation of the trade-offs involved in \ac{jitr} (\autoref{sec:eval}).
\end{itemize}

\section{Just-in-Time Model Replacement}
\label{sec:vision}

\begin{figure}[t]
    \centering
    \includegraphics[width=\columnwidth]{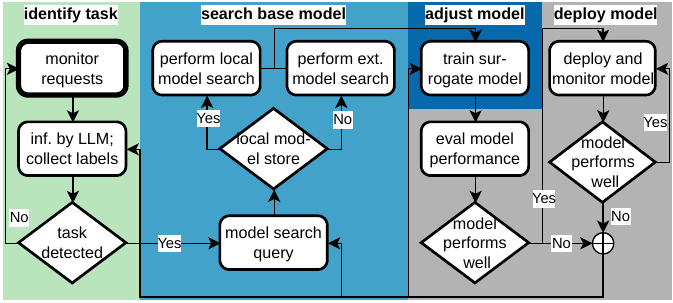}
    \caption{\textcolor{blue}{The \acs{jitr} workflow. The system (i) monitors requests to detect a recurring task, (ii) searches for a suitable base model, (iii) fine-tunes the model to develop a \acl{surrogatemodel}, and (iv) deploys and monitors the \acl{surrogatemodel}, returning to earlier stages if performance is insufficient.}}
    \label{fig:flow-chart}
\end{figure}

In this section, we discuss challenges, solutions, and open research questions in developing a \acf{jitr} system, covering all steps shown in \autoref{fig:flow-chart}. We focus on an efficient model store supporting \ac{jitr} through fast model search since it represents the most critical bottleneck, and the area where the data management community is uniquely positioned to contribute.

We envision several alternatives of user interaction ranging from (i) \emph{no user involvement:} the system switches to \acp{surrogatemodel} without informing or involving the user to (ii) \emph{full user control:} the user has to explicitly approve the use of a \ac{surrogatemodel} and what action is taken when a \ac{surrogatemodel} exhibits performance degradation.  A \ac{jitr} framework may be employed either by the \ac{llm} provider
or by the client transparent to the \ac{llm} provider.

\subsection{Task Identification and Dataset Generation}
\label{sec:vision-collect-data}

As shown in \autoref{fig:flow-chart}, assuming an \ac{llm} is currently used to handle a \ac{rectask}, the first step in developing a \ac{surrogatemodel} is to identify the \ac{rectask} by monitoring and saving incoming requests and responses.
\color{blue}

The input of task detection is the continuous stream of requests posted to the \ac{llm}. This stream may mix  requests that are not part of a \ac{rectask} and requests that are part of a \ac{rectask}. Note that it is not necessarily the case that all requests belonging to a \ac{rectask} use the same prompt template, e.g.,  different versions of an application may use different templates.
For example, for the sentiment classification task some requests may use a template ``What is the sentiment of the following movie review:'' while others use ``Given the following review, what is the user's sentiment?''.

The output of task detection is a set of recurring tasks, where each \ac{rectask} can be represented as \textcolor{red}{one or more \emph{template prompts}} for the \ac{llm}, \textcolor{red}{each of }which is \emph{instantiated} into a concrete prompt by binding values to the template's parameters. In addition, a \ac{rectask} is associated with task metadata (e.g., input/output modalities and task type), performance metadata (e.g., request frequency, latency, and throughput), and a task dataset consisting of incoming requests paired with corresponding responses or ground-truth labels when available.
\color{black}

\paragraph{Challenges}
The main challenge in this step is to detect the \ac{rectask} as early as possible \textcolor{blue}{despite the possible variations in prompts} and gather associated data and metadata without significant overhead during \ac{llm} inference.

\paragraph{Possible Solutions}
Possible approaches for task detection range from user-provided task information to fully automatic task detection.
The easiest approach from a systems perspective is to let the user provide labeled data examples, a prompt template, the task metadata, and minimum performance requirements in a structured format.
Lightweight approaches include extracting a task template for \acp{rectask} from semantic SQL operators,
extend common \ac{llm} application frameworks such as LangChain~\cite{langchain} to allow users to specify \ac{rectask},
or to expect users to describe \acp{rectask} in natural language from which we extract all relevant task information.
\color{blue}
To automatically detect a \ac{rectask}, we can continuously monitor user inputs and analyze their plain text for commonalities,
create task embeddings for all incoming requests,
or monitor the \ac{llm} inference engine's KV cache for repetitive patterns, which could indicate shared prompt prefixes.
\color{black}
A more explicit way to detect a \ac{rectask} and extract metadata is to wrap a subset of incoming requests in an additional prompt that instructs the \ac{llm} to simultaneously answer the request and extract task information.
We can do this online to optimize for costs or offline to optimize for latency.
These methods can also be combined.

\paragraph{Open Research Questions}
A key research question is how to balance task metadata quality against detection overhead.
The higher the metadata quality, the better our decisions on the model development-versus-inference resource-savings trade-off.
For example, knowing up front how often a task will recur and what the performance requirements are enables strategic resource allocation.
Obtaining high-quality metadata requires either greater user effort or a more elaborate detection technique, resulting in higher computational cost.
Another research question is when a sufficient number of requests has been observed to start \ac{surrogatemodel} development.
Earlier triggering increases inference cost reduction by replacing the \ac{llm} sooner, but increases the risk of deploying a \ac{surrogatemodel} with insufficient accuracy, wasting development resources.

\subsection{Base Model Search}
\label{sec:vision-model-store-search}
Once a \ac{rectask} is detected, \ac{jitr} replaces the \ac{llm} with a \ac{surrogatemodel} that minimizes resource consumption while meeting accuracy constraints.
For \ac{jitr} to be effective, we need to minimize the cost and time to monitor the requests, collect training data, and train the \ac{surrogatemodel}, which often makes training from scratch infeasible.
Instead, as shown in \autoref{fig:flow-chart}, we propose to collect training data, use the data to find an appropriate base model using model search~\cite{renggli_shift_2022, alsatian}, and optionally fine-tune this model.
Models may either come from a private model store, if available, or a public model store such as HuggingFace~\cite{hugging_face_hugging_2025}.

The naive approach to model search is to let a human expert select a model from a model store, which requires expert knowledge, often leads to suboptimal choices~\cite{renggli_which_2022}, and is expensive.
Several model search systems have been developed to automate this process~\cite{renggli_which_2022, huang_frustratingly_2022, li_guided_2023}. To reduce the cost of model search relative to the brute force baseline of exhaustively fine-tuning all available models~\cite{renggli_shift_2022}, these systems rank and prune base models based on so-called proxy scores that are meant to approximate the post-fine-tuning performance of a model. For example, a common approach is to
use a model's feature extractor (all layers except the final classification layer) to embed the input data into the model's feature space and train a proxy model (e.g., a fully connected layer) on these features to estimate the task fit, yielding a proxy score (the accuracy of the proxy model on the task data) as an estimate of final model performance. Another common alternative is to use training-free transferability scores such as LEEP~\cite{nguyen_leep_2020}.

\paragraph{Challenges}
While approximate approaches scale better than fine-tuning all models, searching over hundreds of models based on extracted features still takes hours to days~\cite{renggli_shift_2022} and, when applied to thousands or millions of models, search can become more time-consuming than developing a single model~\cite{alsatian, renggli_shift_2022}.
Thus, reducing the time for model search remains the greatest challenge for \ac{jitr}.

\begin{figure}[t]
    \centering
    \includegraphics[width=1\columnwidth]{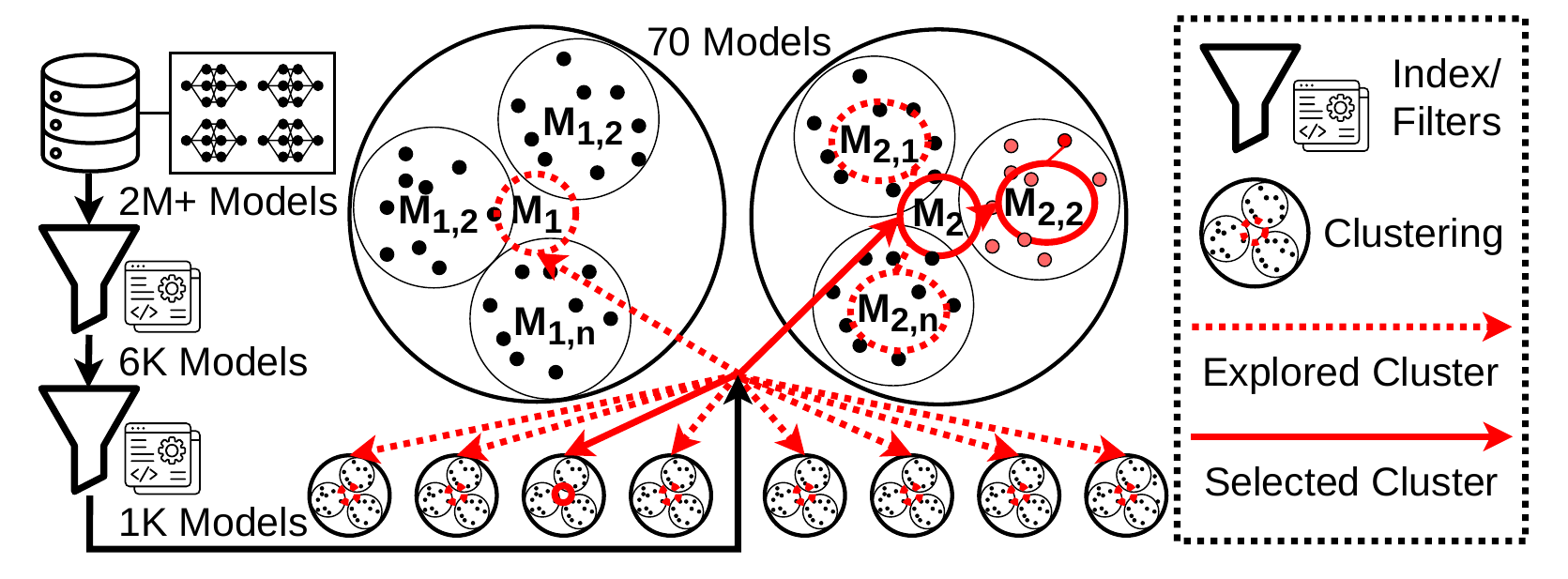}
    \caption{\textcolor{blue}{
    Approximate hierarchical model search. Starting from over 2M models, metadata-based filtering narrows the pool (e.g., to 1,000 models), which is then clustered hierarchically. Promising clusters are explored further (solid arrows); non-promising ones are pruned (dotted arrows), minimizing the models we explore extensively (e.g., to 70 models).
    }}
    \label{fig:advanced-search}
\end{figure}

\paragraph{Possible Solutions}

\color{blue}
The most expensive step in feature-based model search is feature extraction (i.e., loading the model to the GPU and performing inference)~\cite{alsatian}.
There are two main directions for scaling model search, both requiring a co-designed model store:
(i) minimize the number of models considered for feature-based search,
and (ii) reduce model loading and inference times.

\color{red}
Metadata alone is a poor predictor of a model's suitability for a task, but it can serve as a coarse filter, discarding models that are clearly unsuitable, e.g., due to a mismatched modality, poor performance on common benchmark datasets, or their size. Beyond this, many remaining candidates are likely fine-tuned variants of a small number of popular base models or trained on similar data, and can thus be expected to cluster into groups of models with similar performance. This can be exploited by clustering models and evaluating only the cluster representatives with the expensive feature-based search, rather than evaluating every candidate individually.
\color{black}

To minimize the number of models considered for expensive feature-based search, we may first reduce the search space using model metadata, filtering by architecture type, inference latency, memory consumption, and benchmark accuracy on previously seen datasets.
We can also hierarchically cluster models on, for example, task-to-vec embeddings, intermediate inference results, and other metadata~\cite{pal2025modellakes, achille_task2vec_2019}, allowing us to evaluate models in multiple rounds: starting with cluster representatives, pruning non-promising ones, and exploring promising clusters further.
This requires the underlying model store to index models for advanced filtering and support efficient hierarchical clustering.

To reduce feature extraction time, we may prune models on a subset of the data~\cite{renggli_shift_2022} and load approximated model versions (e.g., quantized or pruned) in early exploration stages to identify the most promising clusters, progressively increasing precision in later stages.
We might also exploit if models share identical layers, for example, because they have just been partially fine-tuned by maximizing caching of partial models and intermediate results~\cite{strassenburg_efficiently_2022, alsatian}.
A model store should therefore enable block-wise model access and parameter deduplication as well as fast model access via compressed or approximate model formats~\cite{miao_towards_2017, xiang2025neurstore}.

\autoref{fig:advanced-search} illustrates the search procedure based on hierarchical model clusters.
\color{black}
Starting from a model pool of over 2M models, we filter to a set of 1000 models across multiple stages and cluster them into three levels of 10, 100, and 1000 models, respectively.
At level 0, we evaluate ten models ($M_1, \ldots, M_{10}$) and select the two clusters associated with $M_2$ and $M_5$ as the most promising ones.
On level 1, we evaluate ten models per selected cluster (e.g., $M_{2,1}, \ldots, M_{2,10}$) and again select the two most promising clusters, leading to an evaluation of  40 models at level 2, which totals 70 models for the entire search.

\paragraph{Open Research Questions -- Metadata}
A key research direction is to investigate which metadata types are most valuable for model search and what trade-offs arise in deciding which to store. Models can be annotated with a variety of metadata, such as training data or statistics thereof, the base model, data preprocessing pipeline details, input shape, task type, loss curves, number of parameters, model size, inference latency across hardware platforms, and accuracy on benchmarking datasets~\cite{pal2025modellakes}.
However, not all metadata is effective for model search, and metadata types differ in the effort required to extract and store them~\cite{vartak_mistique_2018, zhang_metastore_2024}. For example, the number of model parameters can be determined automatically to effectively filter large collections of models. Storing intermediate results produced during model inference on reference datasets helps with clustering and speeds up model search, but comes at a high storage cost. Inference latency on unseen hardware is difficult to estimate, yet crucial for model selection.
Finally, it is also important to evaluate how to efficiently store and index all metadata so that it can be rapidly accessed during model search.

Ultimately, we expect a mix of metadata to be most effective.
Lightweight metadata, such as input shape and task type, are easy to extract, small in size, and static, which makes them well-suited for pre-filtering.
Richer metadata, such as intermediate inference results, are more discriminative but expensive to extract and store; limiting it to a representative subset of models can bound both cost and storage.

\paragraph{Open Research Questions -- Model Indexing and Clustering}
Another research direction is model indexing and clustering~\cite{pal2025modellakes}. Open questions include identifying which features are most effective for measuring model similarity and determining which clustering algorithms perform best for organizing models. A key challenge is to assess whether a single, universal clustering scheme can serve all model search queries, or whether multiple specialized clusters or indices are required.
A resulting research question is how to evaluate cluster suitability. Potential approaches include: selecting a representative model for testing,
generating a new model to approximate the entire cluster,
or to try and estimate the range of possible predictions for multiple models at once~\cite{miao_towards_2017}.

\paragraph{Open Research Questions -- Fast Model Access}

Performing inference requires loading each model into memory and can become a major bottleneck~\cite{alsatian}.
Compact storage formats and accelerating model loading, therefore, represent an important research direction.
Key questions include whether searching over approximated models results in the same model ranking or similar proxy scores compared to those from full-precision models
and how storage formats can be optimized for faster loading. Exploring storage designs that are aware of hardware and cloud-specific characteristics could substantially improve model retrieval efficiency.

\paragraph{Open Research Questions -- Multi Tenancy}

Approaches like Alsatian~\cite{alsatian} show that deduplicating computations and model loading within a single model search query yields significant speedups.
At the scale of modern \ac{llm} providers, multi-query optimization to detect shared computations or overlapping data access patterns across queries is a promising direction to improve system efficiency and throughput.

Another direction involves tracking past search results and success rates. For new requests, this enables prioritizing models that have demonstrated strong performance on similar tasks and pruning historically underperforming models early.

\subsection{Surrogate Model Development}
\label{sec:vision-model-development}

Once model search identifies a promising base model, we evaluate whether it performs well enough without further adjustments, or adapt it for the specific \ac{rectask}.
The main challenge is to minimize adaptation time and the time until \ac{surrogatemodel} deployment.
The most promising approach is to fine-tune the model returned from model search using the labeled data generated by the \ac{llm},
distilling the  knowledge~\cite{hinton2015distillingknowledgeneuralnetwork} from the \ac{llm} into the \ac{surrogatemodel}.

\paragraph{Open Research Questions}
Several open questions remain regarding the optimal model development strategy. First, we need to determine what types of knowledge distillation to employ and whether we should extract more than just the final predictions from the \ac{llm}, which would add overhead during \ac{llm} inference.
Second, it is unclear how to predict how much training data is needed to achieve robust model performance. Third, we must investigate to what extent fine-tuning or distillation can compensate for suboptimal choices made during model search.
Finally, there is the question of whether we should apply additional model compression techniques beyond distillation to further improve efficiency.

\subsection{Model Evaluation and Monitoring}
\label{sec:vision-model-monitoring}

Once we have generated a \ac{surrogatemodel}, the question is whether or not to replace the \ac{llm}.
As shown in \autoref{fig:flow-chart}, the replacement process involves multiple stages: identifying whether the \ac{surrogatemodel} is a suitable candidate, deploying it if appropriate, and continuously monitoring its performance to ensure consistent quality. If the \ac{surrogatemodel} proves unsuitable or its performance degrades during deployment, we must adjust our approach by returning to earlier stages such as collecting more labeled data, conducting model search, or performing model fine-tuning.

\paragraph{Challenges}
The main challenges we face are:
(i) How do we validate that a \ac{surrogatemodel} is a good candidate to replace the \ac{llm}?
(ii) Once deployed, how do we ensure consistently high performance, even in the event of distribution shift?
(iii) How do we choose the right tradeoff between competing constraints on cost, inference latency, throughput, and accuracy?
(iv) Which earlier stages of the \ac{jitr} should be repeated when adjustments are needed?

\paragraph{Possible Solutions}

A possible solution to challenge (i) is to collect a validation dataset during task identification, model search, and model development to evaluate the \ac{surrogatemodel}'s accuracy, requiring only a short benchmarking period to obtain performance metrics.
A different solution is to deploy the \ac{surrogatemodel} alongside the \ac{llm} for a certain period of time to compare the \ac{surrogatemodel}'s accuracy and other performance metrics to those of the \ac{llm}.
For challenge (ii), we can also take the approach of parallel deployment, regularly routing a fraction of requests to both the \ac{llm} and \ac{surrogatemodel} to compare their performance.
We can also allow the user to add additional labeled training examples at any point after \ac{surrogatemodel} deployment. The monitor can then employ distribution shift detection~\cite{rabanser2019failing} to determine whether the model needs to be retrained.
For challenge (iii), one solution is
attempting to match the \ac{llm}'s accuracy within a narrow range at all costs while compromising on cost, inference latency, and throughput reduction.
Over time, we can incrementally widen the accuracy threshold and focus more on performance metrics.

\paragraph{Open Research Questions}
Several open research questions remain regarding the optimal monitoring strategy and performance tradeoffs.
We need to determine the right monitoring level and amount of data to collect since unnecessary monitoring introduces overhead.
We also need to establish how to prioritize the different metrics: inference latency, memory consumption, throughput, and accuracy. Equally important is maintaining consistent predictive
performance, since customers are sensitive to fluctuations in prediction quality, which raises the question of how frequently we can switch models without negatively impacting user experience.
Similar to the trade-off for \ac{surrogatemodel} development, decreasing the length of the trial period for a model and frequency of monitoring during deployment increases the risk of failing to detect a poorly performing model, but results in larger cost savings.

\section{\sysname: A Just-in-Time Model Replacement Prototype}

\label{sec:poodle}
\begin{figure}[t]
    \centering
    \includegraphics[width=0.85\columnwidth]{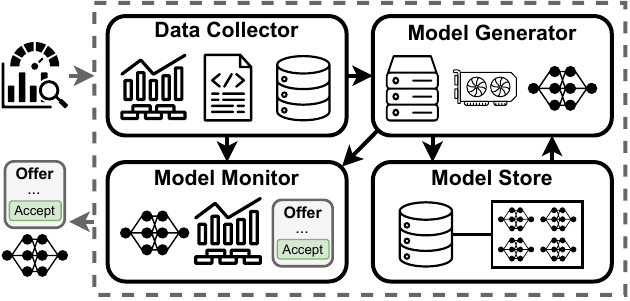}
    \caption{\Acl{jitr} architecture.}
    \label{fig:architecture}
\end{figure}

In this section, we present \sysname, a proof-of-concept prototype for \ac{jitr}.
As shown in \autoref{fig:architecture}, \sysname consists of four components.

The \emph{Data Collector} 
continuously monitors incoming requests and the \ac{llm}'s responses to collect a dataset for training. Additionally, it gathers metadata and performance metrics for request serving, including cost, inference latency, and the number of tokens.
For all additional \ac{rectask} metadata, such as the input data type, the task type, and other metrics, \sysname groups incoming requests by their prefix and, for a subset of requests, uses an additional prompt to simultaneously extract metadata and answer the user query. 

The \emph{Model Generator} receives task definitions from the Data Collector, including the training dataset collected for the task.
It uses a subset of the target dataset to issue a model search request to the \emph{Model Store}.
The Model Store follows Alsatian's~\cite {alsatian} baseline and ranks all models by approximating how well a specific model will perform when being fully fine-tuned.
The Model Generator uses the training dataset consisting of \ac{llm} requests and responses to distill~\cite{hinton2015distillingknowledgeneuralnetwork} the knowledge of the \ac{llm} into the small model via fine-tuning to create a \ac{surrogatemodel}. 
Finally, it persists the \ac{surrogatemodel} in the Model Store for future use.

The \emph{Model Monitor} evaluates candidate models against the deployed \ac{llm}. Utilizing and extending the data collected by the Data Collector, the Model Monitor tracks inference latency, accuracy, and costs for both models.
Once a candidate demonstrates comparable accuracy at a lower cost, the monitor automatically replaces the model, optionally asking the user for approval.

After deployment, the Model Monitor continuously compares the \ac{llm}'s and the new model’s performance to detect data drift.
While we leave the implementation of calculating energy consumption or the CO$_2$ footprint, as well as continuous monitoring, for future work, we estimate the cost savings as follows.
For the \ac{llm} cost, we track the number of input and output tokens and multiply them by the cost as listed by the \ac{llm} provider.
To estimate \sysname's cost, we sum
(i) the cost to process the first $i$ wrapped requests, 
(ii) an estimate of the \ac{surrogatemodel} development cost,
and (iii) the cost for processing remaining requests with the \ac{surrogatemodel}.

\section{Preliminary Results}
\label{sec:eval}

We conduct preliminary experiments to demonstrate the effectiveness of \ac{jitr}.
We examine cost reduction in \autoref{sec:cost}, inference time reduction in \autoref{sec:latency}, analyze the effect of \ac{jitr} on accuracy in \autoref{sec:accuracy}, and show that model search followed by fine-tuning outperforms other model development approaches in \autoref{sec:search}.
Unless stated otherwise, we perform sentiment classification on the IMDB dataset~\cite{maas_learning_2011} and assume that every request processed by the \ac{llm} is wrapped in an additional prompt for metadata extraction, resulting in conservative cost and latency savings estimates.

\subsection{Monetary Cost}
\label{sec:cost}
To determine the cost break-even point, i.e., the number of requests after which \ac{jitr} is cheaper than using an \ac{llm}, we use the prices from \autoref{tab:api-costs}, \textcolor{blue}{and assume a switch from the \ac{llm} to a custom \emph{BERT} model after 5k requests. We further assign a development cost of \$4, which roughly equals three hours of an AWS A10G GPU instance and is a generous budget to search through 30 BERT models and fully fine-tune the top candidate~\cite{alsatian}.}

\begin{table}[t]
\centering
\caption{Model pricing per 1M tokens, April 2026. Models marked with $\dagger$ show TogetherAI prices from August 2025.}
\label{tab:api-costs}
\begin{tabular}{lccc}
\toprule
\textbf{Model} & \textbf{Input (\$)} & \textbf{Output (\$)} & \textbf{Ref} \\
\midrule
GPT-5.5       & 5.00 & 30.00  & \cite{openai_pricing} \\
GPT-5.4-mini  & 0.75 & 4.50  & \cite{openai_pricing} \\
Gemini 2.5 Flash      & 0.30 & 2.50 & \cite{googlecloud2025geminipricing} \\
Llama 3 8B$^{\dagger}$             & 0.20 & 0.20 & \cite{together_llama-3-8b} \\
BERT 80M$^{\dagger}$               & 0.01 & 0.01 & \cite{togethercomputer_m2-bert-80M-32k-retrieval} \\
\bottomrule
\end{tabular}
\end{table}

\begin{figure}[t]
    \centering
    \begin{subfigure}[t]{0.32\columnwidth}
        \centering
        \includegraphics[width=\textwidth]{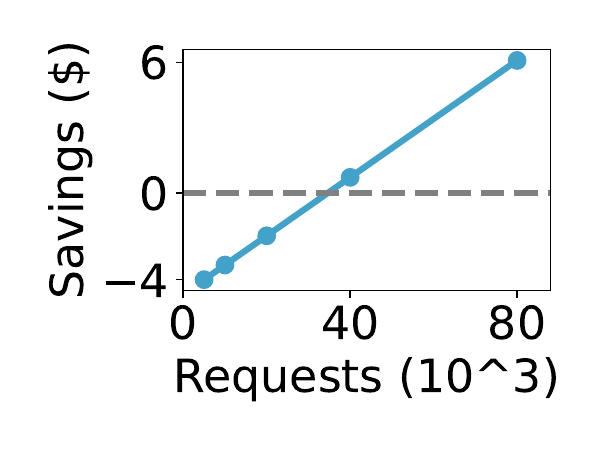}
        \caption{Gemini 2.5 Flash}
        \label{fig:cost-break-even-nano}
    \end{subfigure}
    \hfill
    \begin{subfigure}[t]{0.32\columnwidth}
        \centering
        \includegraphics[width=\textwidth]{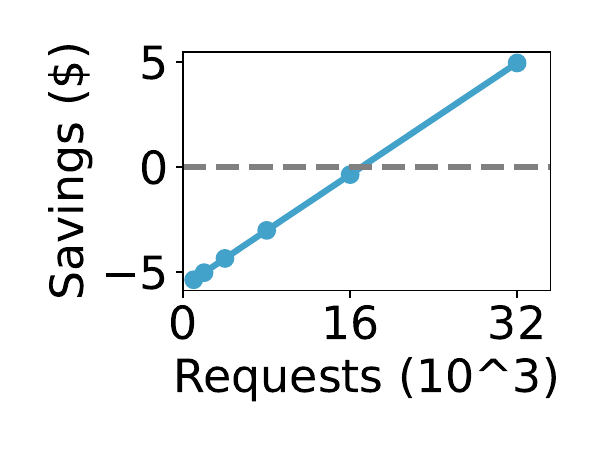}
        \caption{GPT-5.4 mini}
        \label{fig:cost-break-even-gpt}
    \end{subfigure}
    \hfill
    \begin{subfigure}[t]{0.32\columnwidth}
        \centering
        \includegraphics[width=\textwidth]{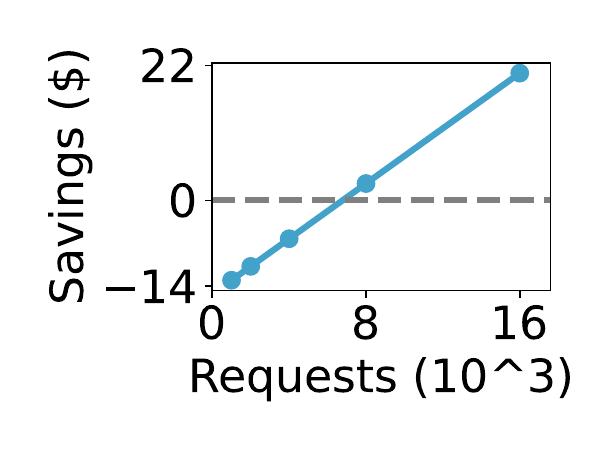}
        \caption{GPT 5.5}
        \label{fig:cost-break-even-llama-405}
    \end{subfigure}
    \caption{\textcolor{blue}{
    Cost break-even analysis for three \acp{llm}.
    }}
    \label{fig:cost-break-even}
\end{figure}

\autoref{fig:cost-break-even} shows the break-even point and cost reduction for different models.
For Gemini 2.5 Flash (used in SemBench~\cite{lao2025sembench}) \sysname breaks even after roughly 40K requests, and reduces the costs from \$141 to \$11 for 1M requests. With larger models such as GPT-5.4 mini, the break-even point is 16K, and for GPT 5.5, the costs break even after 8K requests, with more than \$2,230 saved for 1M requests.

\begin{takeaway}
  Under conservative assumptions for task identification, \ac{jitr} amortizes its overhead after a moderate number of instances and achieves significant cost reductions at scale.
\end{takeaway}

\subsection{Inference Time}
\label{sec:latency}
We determine the inference time break-even point, which is the number of requests after which \ac{jitr} with \sysname becomes faster than using an \ac{llm}.
We use the IMDB dataset and evaluate different batch sizes on an RTX A5000 GPU for
(i) a BERT model,
(ii) a Llama-2-7B model using a base prompt,
and (iii) a Llama-2-7B model using a longer (wrapped) prompt.
An \ac{llm} approach using Llama-2-7B and a \ac{jitr} approach (starting with Llama-2-7B and switching to BERT after 5,000 requests) break even within the first 100K requests.
Llama-2-7B processes 13 items per second at a maximum batch size of 16 while BERT processes $19.6\times$ more items in the same time at a maximum batch size of 128. The time reduction increases with larger request volumes.
For 1M requests, \sysname takes $7.5\times$ less time than the \ac{llm}, and for 2M requests more than $10\times$.
Since Llama-2-7B is a small model compared to a flagship \acp{llm} and all requests are wrapped, we expect real-world speedups to be even larger.

\begin{takeaway}
For local deployment, even when using a small \ac{llm} as the baseline, \ac{jitr} significantly reduces inference time.
\end{takeaway}

\subsection{Accuracy}
\label{sec:accuracy}
In this section, we show that small models can compete with LLMs on simple tasks.
We use 10,000 random items from the IMDB dataset with a 50/50 training-test split and compare the accuracy of an \ac{llm}~\cite{llama2_7b_chat_hf} with the accuracy of a \emph{BERT}~\cite{bert_base_uncased} model fine-tuned on the ground truth or \ac{llm} generated labels.
The \ac{llm} reaches an accuracy of 0.926 on the training and 0.937 on the test data.
The results for the \emph{BERT} model in \autoref{tab:gt_llm_accuracy} show that a \ac{surrogatemodel} reaches competitive accuracy, even though
(i) we do not tune hyperparameters,
(ii) the \ac{llm} might have seen the IMDB data before,
(iii) and that \emph{BERT} uses 256 tokens while the LLM sees 1024.

\begin{table}[t]
\centering
\caption{\textcolor{blue}{
Test accuracy and epochs-to-convergence of a BERT surrogate fine-tuned on IMDB, trained on varying amounts of ground-truth vs. \ac{llm}-generated labels.
}}
\label{tab:gt_llm_accuracy}
\begin{tabular}{rcccc}
\hline
& \multicolumn{2}{c}{\textbf{Ground Truth}} & \multicolumn{2}{c}{\textbf{LLM-Generated}} \\
\cmidrule(lr){2-3} \cmidrule(lr){4-5}
\textbf{\# Items} & \textbf{Accuracy} & \textbf{Epochs} & \textbf{Accuracy} & \textbf{Epochs} \\
\hline
500  & 0.86 & 9 & 0.88 & 6 \\
1,000 & 0.88 & 5 & 0.88 & 7 \\
2,000 & 0.89 & 3 & 0.88 & 5 \\
5,000 & 0.90 & 5 & 0.90 & 2 \\
\hline
\end{tabular}
\end{table}

Related work has come to similar conclusions.
Across multiple studies and datasets, fine-tuned models (e.g., \emph{RoBERTa}) consistently outperform \acp{llm}
on text classification tasks~\cite{bucher2024fine, bosley2023we, edwards2024languagemodelstextclassification, dai2024tailwiz}.
Pangakis et al.~\cite{pangakis2024knowledgedistillationautomatedannotation} avoid test-set contamination and compare \emph{GPT-4} few-shot performance with small models
trained on either human or \emph{GPT-4}–generated labels.
Small models trained on the ground-truth slightly outperform \emph{GPT-4}, and those trained on \emph{GPT-4} labels achieve accuracy nearly matching \emph{GPT-4} itself.

\begin{takeaway}
For well-scoped tasks,
smaller, specialized models achieve accuracy that is competitive with \acp{llm}.
\end{takeaway}

\subsection{Model Search}
\label{sec:search}
In this section, we
show that (1) model search identifies the most promising model and (2) using fine-tuning, outperforms other approaches in development time, accuracy, and required dataset size.

We select ten models from Hugging Face, including the base version of \emph{BERT}, three models trained for sentiment classification ~\cite{socher-etal-2013-recursive, zhang_dive_2023},
and six task-specific non-sentiment classification models to evaluate the rankings produced by model search.
We use Alsatian's~\cite{alsatian} baseline approach and 500 test and training samples to search the models and generate the ground truth ranking by fully fine-tuning all ten models.
Model search ranks domain-specific models lowest, the BERT base model in the middle, and sentiment classification models highest. The highest-ranked model achieves the highest accuracy when being fully fine-tuned.

\begin{figure}[t]
    \centering
    \includegraphics[width=0.97\columnwidth]{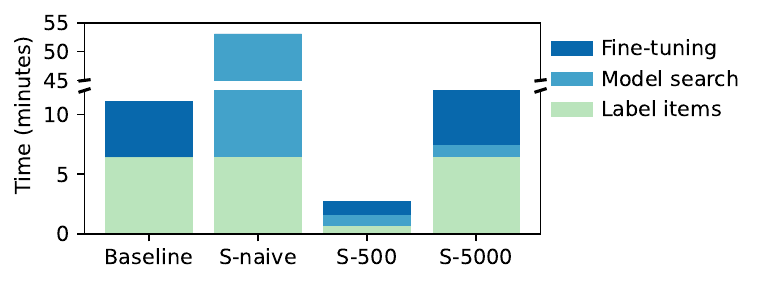}
    \caption{
    Development time breakdown for four surrogate development strategies, all targeting 0.89 accuracy on IMDB. \emph{Baseline} fine-tunes a fixed BERT model; \emph{S-naive} fine-tunes all ten candidate models and then selects the top-performing fine-tuned model; S-500 and S-5000 use model search to select a candidate model by calculating proxy scores for each candidate model and then fine-tune the model returned by model search on 500 or 5,000 samples, respectively.}
    \label{fig:model-dev}
\end{figure}

\textcolor{blue}{In \autoref{fig:model-dev}, we} compare four approaches, all utilizing \ac{llm}-generated labels, with the goal of reaching 0.89 accuracy.
Our baseline collects $n$ \ac{llm}-generated labels and then fine-tunes a standard BERT model to evaluate model development time. For this, a minimum of 5,000 items are required to reach the target accuracy of 0.89, which takes eleven minutes.
The naive search baseline (\textit{S-naive}) takes 53 minutes to fine-tune all ten available models on 5,000 items and select the best-performing one. It achieves an accuracy of 0.92.
The model search approach \textit{S-500} \textcolor{blue}{uses Alsatian's~\cite{alsatian} baseline approach to select} the best-performing model and fine-tunes it on 500 training samples. With under three minutes, it completes 4$\times$ faster than the baseline and 19$\times$ faster than the naive search while reaching an accuracy of 0.91.
With 500 samples for model search and 5,000 samples for fine-tuning, \textit{S-5000} takes twelve minutes and matches the accuracy of the naive search baseline while being $4.4\times$ faster.
When aiming for the highest accuracy using a naive search approach, model search is the bottleneck. In our experiment, this shifts when using a more advanced search method or more data for fine-tuning than for searching. However, the bottleneck shifts back to model search when searching through all 6,000 fine-tuned BERT variants or all 2M models on Hugging Face.
We can mitigate this by a model store that provides system-level optimizations and advanced search capabilities.
Following the approach described in \autoref{sec:vision-model-store-search}, we reduce the effective search space to 70 models. Extrapolating from the \textit{S-5000} setting, which requires approximately one minute to evaluate ten models, this corresponds to an estimated search time of about 7 minutes. By applying additional optimizations, such as partial model access, successive halving~\cite{renggli_shift_2022}, or loading compressed models, the overall search overhead can be further reduced, at which point \ac{llm} inference for data labeling becomes the dominant bottleneck again.

\begin{takeaway}
Model search followed by fine-tuning outperforms alternative model development approaches in development time, accuracy, and required dataset size.
\end{takeaway}

For additional details about Poodle and our preliminary evaluation, see \cite{strassenburg2026poodleseamlesslyscalinglarge}.

\section{Summary and Discussion}
\label{sec:conclusions-outlook}

In this work, we present our vision of \emph{\acf{jitr}} which, transparently to the user, replaces general-purpose \acp{llm} with cheap surrogate models for \acp{rectask}. Our preliminary results demonstrate that \ac{jitr} significantly reduces inference latency and resource requirements for \acp{rectask} while providing the same advantages for these tasks as using an \ac{llm}:
(i) zero manual model development cost,
(ii) no need for manual data collection and labeling,
and (iii) no AI expertise required.
Given the increasing number of publicly available and privately stored models~\cite{delangue_hugging_2023}, it is likely that a suitable model exists that can be fine-tuned to produce an accurate \ac{surrogatemodel}. However, identifying the right base model among millions of models that will lead to low development cost and good accuracy requires further work on improving the performance and effectiveness of model stores and model search.

\begin{acks}
This work was funded by the  German Research Foundation
(ref. 414984028 and ref. 556566056) and NSF awards IIS-2420577 and IIS-2420691.
\end{acks}


\bibliographystyle{ACM-Reference-Format}
\bibliography{sample}

\end{document}